\documentclass{siamonline181217}
\usepackage[T1]{fontenc}
\usepackage{amssymb}
\usepackage{eucal}
\usepackage{textcomp} % textquotesingle not found ??
\usepackage{graphicx}
\usepackage{nicefrac}
\usepackage{float}
\usepackage{todonotes}
\usepackage{color}
\usepackage[latin1]{inputenc}
\usepackage[numbered]{mcode}
\newcommand{\abs}[1]{\left\lvert#1\right\rvert}

\newcommand{\inner}[1]{{\left\langle#1\right\rangle}}

\newcommand{\R} {\mathbb{R}}

\newcommand{\E} {\mathbb{E}}
\newcommand{\V} {\mathbb{V}}

\DeclareMathOperator{\val}{val}
\DeclareMathOperator{\mob}{mob}
\DeclareMathOperator{\mar}{mar}
\newtheorem{Prop}{Proposition}%[chapter]

\newtheorem{Def}[Prop]{Definition}

\begin{document}
\title{Computing Shapley Effects for Sensitivity Analysis\thanks{Version of \today.}}
%\funding{Parts of this work were conducted while the first author was visiting Bocconi University, Milan, Italy.}}}
\author{Elmar Plischke\thanks{Clausthal University of Technology, Clausthal-Zellerfeld, Germany (\email{elmar.plischke@tu-clausthal.de}).} 
\and Giovanni Rabitti\thanks{Bocconi University, Milan, Italy (\email{giovanni.rabitti@unibocconi.it})}
\and Emanuele Borgonovo\thanks{Bocconi University, Milan, Italy (\email{emanuele.borgonovo@unibocconi.it})}}
\headers{Computing Shapley Effects}{Plischke, Rabitti, Borgonovo}
\maketitle
\begin{abstract}
Shapley effects are attracting increasing attention as sensitivity measures. When the value function 
is the conditional variance, they account for the individual and higher order effects of a model input. 
They are also well defined under model input dependence. However, one of the issues associated with their use is computational cost. We present a new algorithm that offers major improvements for the computation of Shapley effects, reducing computational burden by several orders of magnitude (from  $k! \cdot k$  to  $2^k$, where $k$ is the number of inputs) with respect to currently available implementations. The algorithm works in the presence of input dependencies. The algorithm also makes it possible to estimate all generalized (Shapley-Owen) effects for interactions.
\end{abstract}
\begin{keywords}
Shapley effect, Möbius inverse, computer experiments, global sensitivity, pick and freeze sampling
\end{keywords}\section{Introduction}

Computer experiments are widely adopted in modern scientific investigations to simulate natural, 
social and physical phenomena. The increase in computing power allows analysts to develop simulators 
of increasing complexity and analytical approaches fail in delivering insights about the simulator behavior. 
Thus, the input-output relationship is often considered as a black-box.

Sensitivity analysis allows us to shed light on the structure of a black-box model, 
providing the modelers with insights useful in building and interpreting simulator results \cite{SaltTara02}. 
A major concern of computer modelers is the quantification of the relative importance of the inputs of 
the codes. When the model inputs are uncertain, this task is typically performed using global sensitivity analysis methods. 
One can find alternative global sensitivity analysis approaches, including variance-based \cite{SaltTara02}, 
regression-based \cite{StorSwil09} and moment-independent sensitivity methods \cite{Borg07}.

In the computer experiments literature, Shapley values are enjoying an increasing popularity (see, e.g., 
\cite{BallPare18,IoosPrie19,PoolNeme18,RadaSura19} for recent contributions to theory and applications). 
Shapley values originate from game theory \cite{Shap53} and are based on the allocation of the total 
value of a game to the contribution of each player. In the context of sensitivity analysis, Shapley 
values have been proposed for the first time in \cite{Owen14} and called Shapley effects. 
The intuition is to regard model inputs as players and the conditional variance they explain as the value function.  
One main reason for the interest towards Shapley effects is that they remain interpretable also in the 
presence of dependent inputs \cite{SongNels16,OwenPrie17,IoosPrie19} or domain irregularities such 
as holes \cite{OwenPrie17}. In such cases, these importance indices never become negative \cite{OwenPrie17}.

However, the computation of Shapley effects can be very demanding. A seminal proposal can be found in \cite{SongNels16}. 
The work suggests a fourfold algorithm: Looping over all permutations (or a randomly selected subset \cite{CastGome09}), 
looping over the path through a hypercube of index subsets induced by that 
permutation, looping over an outer and 
inner loop for obtaining conditional variances, switching inputs on and off along that path to 
compute the sum of marginal contributions. 
In \cite{BrotBach19}  the algorithm is improved by 
using pick'n'freeze sampling and ignoring high-order terms in presence of block-independent group of variables. 
In \cite{BrotBach18x} an algorithm based on nearest neighbor estimators of the marginal contributions is proposed. 

In this paper, we offer an alternative algorithm.
We proceed in two steps. First, we examine a series of refinements of the current implementation of \cite{SongNels16} 
with the goal of achieving memory and time savings. Then, we change the logic for computing Shapley effects by 
switching from permutations of marginal effects to using M\"obius inverses. 
This restructuring decreases computational cost from   $k! \cdot k$  to  $2^k$ evaluations of a basic sample block, 
where $ k $ is the number of model inputs.

We challenge the algorithm through several experiments, comparing its performance against that of previously 
available algorithms. The experiments show that the proposed implementation leads to the following advantages. 
The algorithm reduces computational burden notably (is memory and time-wise faster than the currently published 
predecessors), leads to unbiased estimates, allows the possibility of computing both Shapley effects and 
Shapley-Owen interaction effects. These latter indices are a generalization of Shapley effects introduced in 
\cite{RabiBorg19x} to study the synergistic/antagonistic nature of interactions among inputs. We remark that 
the algorithms of \cite{SongNels16,BrotBach18x} to compute Shapley effects don't allow one to estimate the 
Shapley-Owen interaction effects. In fact, these two algorithms are based on the permutation representation of 
Shapley effects, which is not available at the moment for the Shapley-Owen effects.

This paper is structured as follows. Section \ref{sec:shapleyvalue} presents the concept of Shapley value from game theory. 
Section \ref{sec:shapleysensitivity} presents the Shapley effects for global sensitivity analysis. 
Section \ref{sec:algorithm} presents an improvement of the algorithm of \cite{SongNels16}. 
The new M\"{o}bius inverse-based algorithm is presented in Section \ref{sec:moebiusalgorithm}. 
Section \ref{sec:experiments} contains the numerical experiments.

\section{Shapley Value}\label{sec:shapleyvalue}
The Shapley value \cite{Shap53} is a concept from cooperative game theory. One considers a game with $k$ players. 
The Shapley value is then the quantity that indicates the worth 
of forming coalitions and the expected payoff for each player.
%\todo[inline]{Note among us: I went on and read Shapley 1953. 
%Shapley 1953 requires not only v(0)=0, but also v to be superadditive (see Shapley 1953 equation 3). 
%Then, is variance superadditive in the presence of dependences? Or has someone after Shapley 
%shown that the superadditivity assumption is, in fact, not necessary? Because I do not know if variance 
%reduction is superadditive under dependent inputs}
Generally, one defines the coalition worth function $\val: 2^{\underline{k}} \to \R_{\geq 0}$ with 
 $\val(\emptyset)=0$, attributing a sum of payoffs to a group of players. 
 Here $2^{\underline{k}}$ is the powerset (set of subsets) of $\underline{k}=\{1,2,\dots,k\}$.

\begin{Def}
Given a coalition worth function $\val$, the marginal contribution of player $i$ joining 
coalition $\alpha$ is $\mar(\alpha,i)=\val(\alpha\cup \{i\})-\val(\alpha)$.

The Shapley value is then defined by
\begin{equation}
\begin{split}
\Phi_i(\val{})&= \sum_{\alpha: i\not\in \alpha} \frac{\abs{\alpha}! (k-\abs{\alpha}-1)!}{k!}\left(\val(\alpha\cup \{i\})-\val(\alpha)\right)\\
&=\frac{1}{k}\sum_{\alpha: i\not\in \alpha} {\binom{k-1}{\abs{\alpha}}}^{-1}\mar(\alpha,i).
\end{split}\label{eq:shapley}
\end{equation}
\end{Def}
\begin{Prop}\label{Prop:Shap}
The Shapley value of player $i$ is characterized by the following four axioms,
\begin{itemize}
\item Pareto-efficiency: $\sum_{i=1}^k \Phi_i(\val{})=\val(\underline{k})$
\item Symmetry: If  $\val(\alpha\cup \{i\}) = \val(\alpha\cup \{j\})$ for all subsets $\alpha$ containing neither $i$ nor $j$ then 
$\Phi_i(\val{})=\Phi_j(\val{})$
\item Linearity: $ \Phi_i(\val_1+\val_2)=\Phi_i(\val_1)+\Phi_i(\val_2)$
\item Null-player: If for all $\alpha$, $ \mar(\alpha,i)=0$ holds then $ \Phi_i(\val)=0$.
\end{itemize}
\end{Prop}

One can interpret the Shapley value either as payoff from joining a coalition or from leaving the anti-coalition.
An alternative route to computation is offered by the formula of \cite{Grab06,Owen14}, for which one needs
 to calculate the
M\"{o}bius inverses $\mob(\alpha)$ of the value functions $\val(\alpha)$. These are defined implicitly by
$\val(\alpha)=\sum_{\beta\subseteq \alpha}\mob(\beta)$ and therefore
$\mob(\alpha)=\sum_{\beta\subseteq \alpha}(-1)^{\abs{\alpha}+\abs{\beta}}\val(\beta)$ \cite[Chap. 8]{Mazu10}, \cite{Rota64}. 
Then
\begin{equation}\label{eq:ShapleyMoebius}
\Phi_i(\val{})= \sum_{\alpha: i\in \alpha} \frac{\mob(\alpha)}{|\alpha|}.
\end{equation}
Hence each M\"{o}bius inverse is weighted by the number of members in the coalition: 
Player $i$ gets full credit for the games won by one-self, half credit for those were the player teamed up in pairs, etc. 

\section{Shapley Effects for Sensitivity Analysis}\label{sec:shapleysensitivity}
In sensitivity analysis, one usually asks the question of the extent with which an 
uncertain input influences the outcome of a complex simulation code \cite{BorgPlis16}.
Hence, the role of the players is taken by the inputs and $k$ is the input 
dimension of the simulation model $g: \R^k\to \R$.
 
For sensitivity analysis, the coalition-worth function 
is taken to be the variance of conditional expectation of $Y$ given $X_i$ or the ratio between this 
quantity and the unconditional variance of $Y$,
$\val(\alpha)=\V[Y]^{-1}\V[\E[Y|X_\alpha]]$ \cite{Owen14,OwenPrie17} which then accounts for a grand total of one.
For these choices of value functions, the term Shapley effects has been coined. 
 The value functions are then equal to the
 subset importance of \cite{LiuOwen06}. Under input independence, the M{\"o}bius inverses coincide with
 the variance-based first order and 
 higher order Sobol' effects, e.g. $\mob(\{i,j\})=\val(\{i,j\})-(\val(\{i\})+\val(\{j\}))$ is the contribution to 
 the output variance stemming from the pairwise interaction/second order effect of inputs $i$ and $j$.
 Also under input independence, as first order effects $S_i$ satisfy $S_i=\val(\{i\})=\mob(\{i\})$ and 
 total effects satisfy
$T_i=\sum_{\alpha: i\in \alpha}\mob(\alpha)$ we have $S_i\leq \Phi_i(\val{})\leq T_i$. So 
the need for computing Shapley effects arises only if the gap between first order and 
total effects is large or dependences in the input are present. As a consequence 
from \eqref{eq:ShapleyMoebius} and Pareto-efficiency we observe the following results.
\begin{Prop}\label{Prop:SobSi}
Under input independence,
the Shapley effect is bounded by the mean of the main effect and the total effect,
$\Phi_i\leq \frac{1}{2}(S_i+T_i)$. If equality holds then there are no interaction 
terms of order larger than two involving input $i$. If the sum of all main and 
total effects equals $2$ then only pairwise interactions may be present in the model.
\end{Prop}

We also observe that \cite{SongNels16} show the following duality result: considering the value function 
$\val'(\alpha)=\E[\V[Y|X_{\sim\alpha}]]$ instead of $\val(\alpha)=\V[\E[Y|X_\alpha]]$ 
leads to the same Shapley effects. Both of these value functions can be estimated from a pick'n'freeze design.

\subsection{Shapley Effects for Groups}
Recently, \cite{RabiBorg19x} introduce a Shapley effect for groups, 
building upon results by \cite{Owen72} and \cite{GrabRoub99}.
The following expression, termed Shapley-Owen effect for the group $\alpha$,  parallels \eqref{eq:ShapleyMoebius},
\begin{equation}\label{eq:ShapleyMoebiusGroup}
\Phi_\alpha(\val{})= \sum_{\beta: \alpha\subseteq\beta} \frac{\mob(\beta)}{|\beta|-|\alpha|+1}.
\end{equation}
Hence, having available the M\"o{}bius inverses allows one to obtain these Shapley-Owen effects.
Note that they are governed by a slightly different set of axioms than in Prop.~\ref{Prop:Shap}.

If all M\"o{}bius inverses are nonnegative, we can generalize the findings of Prop. \ref{Prop:SobSi}, as
\begin{equation}\label{eq:boundShapGroup}
S_\alpha=\mob(\alpha)\leq \Phi_\alpha(\val{})\leq \sum_{\beta: \alpha\subseteq\beta} \mob(\beta)=\Upsilon_\alpha
\end{equation}
where $\Upsilon_\alpha$ is the superset importance measure of \cite{LiuOwen06}.
We can further sharpen the upper bound in 
\eqref{eq:boundShapGroup},
\begin{equation}\label{eq:boundsharpShapGroup}
\Phi_\alpha(\val{})\leq 
\mob(\alpha)  +  \frac{1}{2} \sum_{\beta: \alpha\subsetneq\beta} \mob(\beta)
=\frac{1}{2}\left( \mob(\alpha) 
+\sum_{\beta: \alpha\subseteq\beta} \mob(\beta) \right).%=\frac{1}{2}\left(S_\alpha + \Upsilon_\alpha \right).
\end{equation}
We then formulate an interaction analogon of Proposition \ref{Prop:SobSi}.
\begin{Prop}
	Under input independence,
	the Shapley-Owen effect  for input group $\alpha$ is bounded by the mean of the Sobol' index and the superset importance of $\alpha$, that is
	$\Phi_\alpha\leq \frac{1}{2}(S_\alpha+\Upsilon_\alpha)$. If equality holds then there are no higher order interaction terms larger than $|\alpha|+1$ involving $\alpha$. 
	%If the sum of all main and total effects equals $2$ then only pairwise interactions may be present in the model.
\end{Prop}
%\todo[inline]{Elmar can you check this proposition? It should hold since if there are no interactions then $ \Upsilon_\alpha=S_\alpha$}

\section{A Refinement}\label{sec:algorithm}
A first algorithm we study is based on the sample approach of \cite{SongNels16}, with some marginal modifications aimed at improving estimation accuracy. We propose to:
%\todo[inline]{Improvements over what algorithm?}
\begin{itemize}
\item Use of a pick'n'freeze design \cite{GambJano16,BrotBach19} instead of a brute-force double loop to compute conditional variances 
\item Use of duality result: two estimators can be obtained for the same computational costs
%\todo[inline]{Dear Elmar, what is the advantage here? maybe we need a line of comment on this item below}
\item Estimation of conditional variances via Sobol'/Saltelli and Jansen formulas to take advantage of the pick'n'freeze design and the duality result
%\todo[inline]{Pick and Freeze has it also been done by Broto?}
% \todo[inline]{Does this not overlap with the pick and freeze improvement?}
\item Use of quasi Monte-Carlo (QMC) design for improved convergence \cite{JoeKuo08} compared to a crude Monte-Carlo design
%\todo[inline]{Sorry for my doubt, but are we sure that QMC is always an advantage w.r.t. crude MC when there is pick and freeze?}.
\end{itemize}

%\todo[inline]{The "now" lets one think that "before" they were producing noise, but there is no "before" in this descripton. Maybe we need an example to clarifying what the paragraph is talking about?}
The convergence properties of pick'n'freeze designs are discussed in \cite{GambJano16}. Combined with QMC sampling, one profits of the accelerated convergence of the variance estimates 
for a large class of functions (i.\,e.\ those bounded in the sense of Hardy--Krause).
Moreover, the pick'n'freeze design allows one to identify dummies exactly, 
as theoretically expected (null-player property of Prop.\ \ref{Prop:Shap}),
eliminating numerical noise stemming from the variability of variance estimates of different samples as in \cite{SongNels16}.
The Jansen estimator for superset importance and the Sobol'/Saltelli estimator for subset importance 
offer two different ways of obtaining Shapley effects.

A version ready to test in \textsc{MatLab} or Octave is available in Algorithm \ref{alg}. 
The calling convention requires as parameters the dimension of the model $k$, 
the size of a basic sample block $n$,  %follows the EFAST implementation of \cite{Ekst05}, 
the simulation model and the transformation from the unit hypercube 
into the desired marginal distributions. Both the model function and the input space transformation are assumed to be vectorized.

Further optimizations of Algorithm \ref{alg} are possible. Specifically, the ways 
in which indices of model inputs are selected or deselected by traversing all different permutation paths
%vertices in the graph which is spanned by the selection or deselection of  are
include the same marginal contribution multiple times.
%visited multiple times , hence increasing computational time
%unnecessarily. 
%\todo[inline]{Giovanni, Elmar: Vertices appear for the first time}
Filling a database of coalition 
worth value functions as one traverses these paths and querying the database before a model evaluation
greatly enhances the applicability, especially in higher input dimensions: There are 
$2^k-1$ value functions for a model with input dimension $k$,
as opposed to $k! \cdot k$ evaluations of marginal functions when 
looping over all permutations
%applying \eqref{eq:shapley} directly 
($k! \cdot k\gg 2^k$). 
This requires a database query 
before the model evaluation and an insertion after the computation of a value function. % in lines 6 and 15 of Algorithm \ref{alg}. 
%(Without a distributed computing setup, this might be implemented as a simple lookup from a key-value list).
%
%\todo[inline]{sentence in parenthesis had to follow}

Furthermore, 
the direct storage of all permutations might impair performance (\textsc{MatLab 2018} warns that 
for 11 dimensions the permutation matrix will occupy more than 3 Gb).
Hence, Heap's algorithm \cite{Heap63} may be used to generate the possible permutations iteratively 
instead of filling a large matrix with all combinations.
%The for-loop construction in lines 9 and 24 of Algorithm \ref{alg} can then be replaced by an
% iterative way of obtaining permutations. 

When referring to Algorithm \ref{alg} in the reminder, we shall consider the optimized version which implements
these two improvements and which is available upon request. 

\begin{algorithm}\caption{A {\sc MatLab}/Octave Implementation of Shapley Effects (simplified).}\label{alg}%
\begin{lstlisting}
function [Shap,Shap2,evals]=shapley(k,n,model,trafo)
% SHAPLEY Shapley effects.
u=sobolpoints(n,2*k); % net(sobolset(2*k),n)
xa=trafo(u(:,1:k));xb=trafo(u(:,k+1:end));

ya=model(xa);yb=model(xb);evals=2*n;
Shap=zeros(1,k); Shap2=zeros(1,k);

for p=perms(1:k)' % for each column
val=0; % coalition worth of empty set
val2=0;
xi=xa;
 for q=p'
  xi(:,q)=xb(:,q); % winding stairs
  yi=model(xi);evals=evals+n;
  nval=mean((yi-ya).^2); % superset importance
  Shap(q)=Shap(q)+nval-val; % marginal contribution
  val=nval;
  
  nval2=yb'*(yi-ya); % subset importance
  Shap2(q)=Shap2(q)+nval2-val2;
  val2=nval2; 
 end
end
Shap=Shap/2/var(yb)/prod(1:k)
Shap2=Shap2/n/var(yb)/prod(1:k)
\end{lstlisting}%
\end{algorithm}

Under input dependence, a Rosenblatt transformation \cite{MaraTara12}
%\todo{trasnformation of a dependence? should it be a transformation of the inputs?}  
is needed in order to compute dependent conditional input realizations of a input group given realizations 
of the complementary index group.
This can be effectively implemented in the Gaussian copula case (rank correlation, normal-to-anything-transformation, 
Nataf transformation) using the upper-triangular Cholesky roots of the reordered correlation matrix.
%Line 4 of Algorithm \ref{alg} then contains a code snippet \verb|x=trafo(normcdf(norminv(u)*chol(S)));|
%line 14  has also to be modified in order to sample \verb|xa| conditional to \verb|xb(:,q)|.
One might argue that the quasi Monte-Carlo structure is distorted by the Cholesky root and a 
symmetric matrix root may be more suitable, however, then the triangular structure and hence 
the Rosenblatt transformation property is lost.
%\todo[inline]{Giovanni, Elmar, i would suggest some change in the wording; it gives the impression that this piece of the paper resembles a userguide for the algorithm, but of course we can also leave it like that if we wish to be faster}

In case of input dependence, the Jansen estimator of superset importance (as implemented in Lines 16--18) becomes invalid  
because in this design the input blocks 
for \verb|yi| and \verb|ya| of the complementary index set $\sim\alpha$
 are not independent. However, the code for the Sobol'/Saltelli estimator (in Lines 20--22)
stays valid as \verb|yb| and \verb|ya| are using independent inputs. Hence in the dependent case,
the duality between using $\val{}$ and $\val'{}$ is lost, but only due to the Jansen estimator breaking down. 
Alternative estimators (as discussed below) are still working. 
%\todo[inline]{there is sort of a logical gap between Jansen (we discuss before) and then duality. 
%Maybe we have to say that duality breakes before discussing estimation under dependence?}
In the next section, we discuss an alternative approach based on an intuition that notably reduces 
computational burden. The improved algorithm introduced in this section will then be compared against 
the new algorithm and the original implementation of \cite{SongNels16} in a series of experiments in Section \ref{sec:experiments}.

\section{Alternative Approach Using the M\"{o}bius Inverse}\label{sec:moebiusalgorithm}
\begin{figure}
\centering
\includegraphics[width=0.67\textwidth]{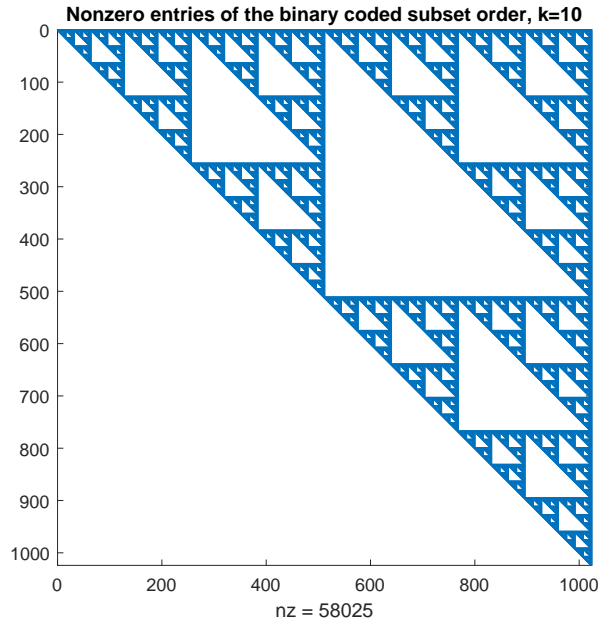}
\caption{The subset order relation is the Sierpinski gasket, Pascal's triangle modulo 2.}\label{fig:sier}
\end{figure}

As the M\"{o}bius inverses used in \eqref{eq:ShapleyMoebius} are formally equivalent to the functional 
ANOVA decomposition terms, \eqref{eq:ShapleyMoebius} 
offers a viable alternative for computing the Shapley effects. Then the $2^k-1$ value functions have to be computed
and a system of equations using a sparsely populated $(2^k-1) \times (2^k-1)$  matrix has to be solved. 
We investigate if this scheme is more attractive than via the computation of all marginals. 
\begin{algorithm}\caption{A {\sc MatLab}/Octave Implementation of Shapley Effects using Möbius Inverses.}\label{alg2}%
\begin{lstlisting}
function [Shap,V]=shapleymoebius(k,n,model,trafo)
% SHAPLEY Shapley effects using Möbius inverse.
u=sobolpoints(n+1,2*k);u(1,:)=[]; 
xa=trafo(u(:,1:k));xb=trafo(u(:,k+1:end));
ya=model(xa);yb=model(xb);

if(k>=log2(flintmax)), warning('Precision (and patience) may be lost.');end
l=2^k-1;H=zeros(2,l);sz=zeros(1,l); for i=1:l
% selection of input subset
 g=bitand(i,2.^(0:k-1))~=0; % lsb codes first index
 sz(i)=sum(g); % subset size
 xi=xa; xi(:,g)=xb(:,g);
 yi=model(xi);
 H(:,i)=[mean((yi-ya).^2)/2;(yb'*(yi-ya))/n];
end
%% Shapley effects via Möbius Trafo: 
% poset inclusion matrix is Pascal triangle mod 2
mob=zeros(2,l); sel=1; for i=1:l
    ii=find(sel);
    mob(:,i)=(H(:,ii)*(-1).^(sz(i)+sz(ii)'))/sz(i);
    sel=xor([1,sel],[sel,0]);
end
%% Owen/Grabisch formula (weights are already included)
Shap=ones(2,k); for i=1:k
    Shap(:,i)=sum(mob(:,logical(bitand(1:l,2^(i-1)))),2);
end
%% variance
V=H(:,end);
\end{lstlisting}%
\end{algorithm}
Suppose that all $2^k-1$ non-vanishing subsets are indexed, $\alpha_j,j=1,\dots,2^k-1$ and the 
value functions have been gathered in
$H_j=\val(\alpha_j)$. Let the matrix $Z$ code the partial ordering,
\begin{equation*} 
Z_{jl}=\begin{cases}1 & \text{if}\quad \alpha_j\subseteq\alpha_l\\ 0, & \text{otherwise}.\end{cases}
\end{equation*}
Via this coding, subset inclusion becomes a logical implication between the matrix columns: $\alpha_j$ is a subset of $\alpha_l$ 
if for all $m$ with $Z_{mj}=1$ it follows that $Z_{ml}=1$.
The Möbius inverses are then 
obtained from $M=HZ^{-1}$ 
and by \eqref{eq:ShapleyMoebius}
the Shapley effect for input $i$ is given by $\Phi_i=\sum_{j: i\in\alpha_j} \frac{M_j}{|\alpha_j|} $. Experiments show that 
this matrix 
has a relative number of non-zero elements  $\left(\frac{3}{4}\right)^k$ compared to the total size of $(2^k-1)^2$.
For larger dimensions, despite being sparsely populated, processing this matrix remains out of reach.
If the set of all subsets is obtained via binary coding (the least significant bit ($2^0$) codes the first index, etc.),
then the matrix $Z$ coding the subset inclusion is Pascal's triangle modulo 2 for which we can construct the columns 
by an exclusive-or operation. The theoretical details for the binary representation of this Sierpinski gasket 
are found in \cite{PeitJurg04}. Figure \ref{fig:sier} shows the subset order relation: A new input $i$ 
is considered at the subset indexed by $2^i-1$ and stays active for $2^i$ subsets.
For inversion we need to determine the signs $(-1)^{|\alpha_i|+|\alpha_j|}$ from
the length of the subsets being compared to one another. The $ |\alpha_i|$ lengths of the multi-indices 
are also needed for the weights in \eqref{eq:ShapleyMoebius}.
Hence we avoid forming and inverting the $Z$ matrix altogether,
see line 20 of Algorithm \ref{alg2}, which is a reference implementation.  As in Algorithm \ref{alg}, pick and freeze sampling of 
quasi Monte-Carlo Sequences as well as both subset and superset importance are used for the computation. 

Again, under dependence the Jansen estimator the first component in line 14 of Algorithm \ref{alg2} is not an estimate of superset importance.
Instead, \verb|yb'*(yb-yi)/n| may be used which estimates the superset importance of the complementary set. The index into the 
complementary set is then obtained by $2^k-1-i$ for $i\neq 2^k-1$. 
  
If all the M\"obius inverses are available then further sensitivity metrics like effective dimensionality may be computed \cite{BallPare18}.
Note that both presented algorithms do not take shortcuts by assuming that higher order contributions vanish or 
by considering a randomized selection of permutations.
If the model evaluation takes virtually no time, models with input dimensions up to $k=20$ are computationally tractable with Algorithm \ref{alg2}.
%\todo[inline]{maybe these two sentences could go in the presentation of the algorithm?}

\section{Numerical Experiments}\label{sec:experiments}
In this section,
we test Algorithms \ref{alg} and \ref{alg2} and compare their performance with the algorithm of \cite{SongNels16}
on benchmark analytical models used in previous works on the estimation of Shapley effects. 
As an application, we consider the fire-spread model studied in \cite{SongNels16}.
For brevity, we shall use the following abbreviations in this section: EP will refer to the algorithm of \cite{SongNels16}, 
based on exact permutations, OP refers to Algorithm \ref{alg}
including database lookup and iterative generation of permutations, and MI is Algorithm \ref{alg2} based on M\"obius inverses.
\subsection{Analytical Models}
Let us consider the Ishigami function
\begin{equation*}
Y=\sin(X_1)\left(1+\frac{1}{10}X_3^4\right)+7\sin(X_2)^2,\qquad X_i\sim \mathcal{U}(-\pi,\pi), \quad i=1,\dots,4.
\end{equation*}
The inputs are uniformly distributed on $(-\pi,\pi)$. In the analysis we included a dummy input, $ X_4 $.
The variance based sensitivity measures of all orders are known for this model \cite{SaltChan00}: 
Because the inputs are independent, using \cite[Theorem 1]{Owen14} we have the corresponding Shapley effects:
$\Phi_1=0.4358$, $\Phi_2=0.4424$, $\Phi_3=0.1218$. Regarding interactions,
the only non-vanishing Shapley-Owen effect is $\Phi_{13}=0.2437$.

To estimate the Shapley effects, in OP and MI we use a base sample from a scrambled Sobol' quasi Monte-Carlo sequence of size 1024.
The EP algorithm uses three runs computing conditional variances in the inner MC loop and 64 for the outer loop to estimate 
the mean of conditional variances, while
 the unconditional output variance is estimated from a sample of size 1024. 
 
%Results show a comparable performance of Algorithms \ref{alg} and \ref{alg2}. 
%With a basic sample of size $1024$ the error is below $2\%$,
%and for a sample of size $4096$ the error is vanishing. 
% \todo[inline]{This cannot be seen from the figure, which is all at 1024. Anywhere else in the paper? Otherwise, I would eliminate and go directly to the comments of Figure 6.1}
 % at all sample sizes. 
%\todo[inline]{This is not clear, Figure 6.1 shows that there is some dispersion}

\begin{figure}[htp]
\centering
\includegraphics[width=0.9\textwidth]{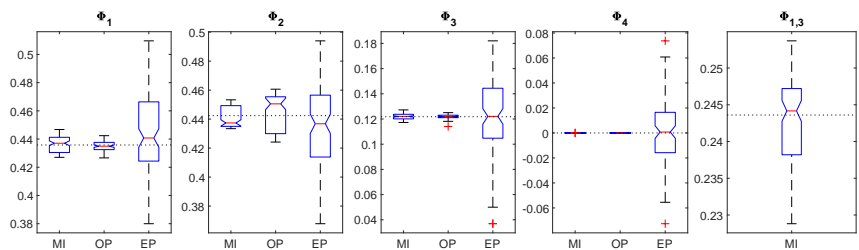}
\caption{Shapley effects for the Ishigami function. Comparison of MI, OP and EP algorithms. 
Rightmost graph: Shapley-Owen effect for the pair $(X_1,X_3)$.}\label{fig:Ishi0}
\end{figure}
The results are displayed in Figure \ref{fig:Ishi0}, showing box-plots of 100 replicates.
The analytical values are marked by a dotted line.
MI and OP show a comparable performance. The dummy parameter is identified correctly by both algorithms.
%\todo[inline]{maybe we could add what is the star also in the caption; by the way, is the star an average over the replicates or a point estimate? By the way, it canno be "the estimate of MI", but, maybe, "the estimate obtained with the MI algorithm"...}
\begin{table}
\caption{Ishigami test function with dummy parameter. Details on the computation.}\label{tab:Ishi}
\centering
{\small% % Table does not match figure ...
\begin{tabular}{c|c|c|c}
 & M\"obius Inverse & Optimized Permutations & Exact Permutations \\
\hline 
Computational time & 5s & 2s & 2m05s\\
Model runs & $15360$ %(=1024\cdot(2^4-1))$ 
& $16536$% % = 1024*2^4$ 
& $14848$ %(=1024+4!\cdot 3 \cdot 64 \cdot 3)$
\\
Quadratic Risk & $9.84\cdot 10^{-5}$ & $9.05\cdot 10^{-5}$ & 0.0035
\end{tabular}}
\end{table}
In Table \ref{tab:Ishi} further details are reported. The computational time includes all 100 replicates, 
while the evaluation count is for a single run of the code. 
For MI and OP computations where performed in \textsc{MatLab} 2018a, while EP is available from the
\textsc{R} sensitivity package\footnote{A stripped down version of EP takes 6s under \textsc{MatLab}, 
while MI and OP take 50s and EP takes 250s under \textsc{Octave} 4, computing 100 replicates.}. 
A desktop Windows 7 64 
bit machine with i7 processor, 8 Gb RAM was used for the simulations.
The number of model evaluations required by the algorithms under comparison is of the same order. 
The OP algorithm has replaced 82 out to the 98 block model evaluations by a database lookup.
As in \cite{BrotBach18x},
we also report the quadratic estimation error given by 
$\sum_{m=1}^k \mathbb{E}\left[\left(\hat \phi_m -\phi_m\right)^2\right]$ in Table \ref{tab:Ishi}.The quadratic risk
error term is lower for MI and OP, due to an exact zero for the dummy parameter and also due to the use of a QMC design. 
%and the dummy variable is exactly identified as inactive.
The MI algorithm allows one to compute Shapley-Owen effects for groups. We have done so for the only non-vanishing second-order Shapley-Owen effect,
which is reported in the rightmost plot of Figure \ref{fig:Ishi0}.  

\begin{figure}
\centering
\includegraphics[width=0.8\textwidth]{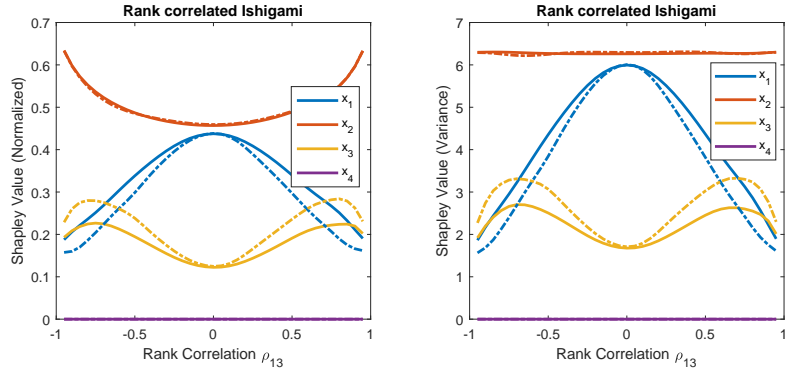}
\caption{Shapley Effects for the rank-correlated Ishigami function.}\label{fig:1}
\end{figure}
For the dependent input case, we introduce a rank correlation between inputs $1$ and $3$ as in \cite{KuchTara12}. 
The results for a basic sample block of 1024 are shown in Figure \ref{fig:1}. The four lines in Figure \ref{fig:1} represent the Shapley effect of each of the inputs as the correlation between $X_1$ and $X_3$ varies from $-1$ to $1$.
The left graph shows the Shapley effects when the value function is the relative variance contribution, i.e. normalized to one, the right graph when the value function is the absolute variance contribution.
The dashed lines are the Shapley effect estimates obtained with the Jansen estimators. As soon as $\rho \neq 0$, these estimates differ from the Sobol'-Saltelli estimates. While the Sobol'-Saltelli estimates are in 
agreement with the output from the algorithm of \cite{SongNels16}, the Jansen's estimates are not.  
This becomes clear observing that the output from the 
sample block chosen conditional to the $\alpha$ 
subset being fixed at positions from the B sample is not the same as the complementary subset $\sim  \alpha$ chosen 
conditionally from the A sample block in the pick'n'freeze estimation scheme under dependence.  

Note that one may infer the wrong conclusions when not keeping in mind that the grand total of the output variance may vary: In absolute terms 
the share attributed to $ X_2 $ does not change, but in relative terms it becomes more important under 
co- or contramonotonic behaviour of $X_1$ and $X_3$.

As a second model, we use Sobol $ g $ functions. These functions have provided useful test cases for variance-based 
sensitivity measures \cite{SaltSobo95,SaltAnno10a,TouzBusb13,JimeGilq18}, but have not been used in the association with Shapley effects yet. 
We test an $8$ dimensional setting using the parameterization of \cite{SaltSobo95}, writing:
$$Y=\prod_{i=1}^8 \frac{|4X_i-2|+a_i}{1+a_i},$$
where $X_i\sim \mathcal{U}[0,1]$ and $a_1=a_2=0$, $ a_3=3$, $ a_4=\cdots =a_8=9$.
For this model, one can obtain the Shapley effects analytically under independence. 
The (unnormalized) value functions are $ \val{(\alpha)} =\V[\E[Y|X_\alpha]]=\prod_{i\in\alpha}(1+V_i)-1$
 where $V_i=\V[\E[Y|X_i]]=\frac{1}{3}(1+a_i)^{-2}$ \cite[Appendix A]{SaltAnno10a}. 
By \cite[Theorem 1]{Owen14}, the analytical normalized Shapley effects are $\Phi_1=\Phi_2=0.469$,   
$\Phi_3=0.0341$,  $\Phi_{4,\dots,8}=0.00551$, the estimated values are $\hat\Phi_1=\hat\Phi_2=0.48$,   
$\hat\Phi_3=0.03$,  $\hat\Phi_{4,\dots,8}<0.01$.
%\todo[inline]{Checjk that these values obtained analytically?}
For this estimation, we used MI and OP algorithms with a block sample size of $n=1024$ using QMC. 
%\todo[inline]{qMC or MC?}
Regarding computational time, OP takes under 10s, while MI takes 0.08s. 
In terms of numerical results, the behavior is similar to the one registered for the Ishigami function: at $ n=1024 $, 
estimates display a negligible numerical error with respect to the analytical values. 
%\todo[inline]{has the optimized version of the algorithm been presented before?}

It is well known that the number of terms involved in the calculation of Shapley effects
increases exponentially with the simulator size. We now consider the 15 dimensional test function presented in \cite{OaklOHag04}, 
\begin{equation*}
Y=g(X)=\inner{a_1,\mathbf{X}}+\inner{a_2,\sin(\mathbf{X})}+\inner{a_3,\cos(\mathbf{X})}+\inner{\mathbf{X},M\mathbf{X}},
\end{equation*}
where inner products are being formed with prescribed vectors $a_1$, $a_2$, $a_3$ and matrix $M$. 
The inputs $\mathbf{X}=(X_1,\dots,X_{15})$ are independently and standard normally distributed.
%\todo[inline]{Sample size and random number generator?}
Using a QMC sample of size $2.048$, the MI algorithm takes around 30s, on the same desktop machine as reported before. 
However, the OP algorithm becomes now time-wise inefficient and does not deliver results in a reasonable amount of time. 
(In the same time span which MI needed to compute the results, the OP algorithm processed less than $1\%$ of all marginals.)

The normalized Shapley effects obtained with the MI algorithm
can be found in Table \ref{tab:1}. The Shapley effects are located between the first order and total effects.
As the model only has only up to second order interactions, theoretically the Shapley effects should be the mean 
between first order and total effects (Proposition \ref{Prop:SobSi}): this can be seen from the sum of both 
the first order and the total effects 
being close to $2$.  
\begin{table}
\caption{Results for the Oakley/O'Hagan test function, basic sample block 2048 QMC. First and total effects 
are extracted from the M\"obius inverses.%$\mob({i})=\mob(\alpha_{2^i-1})$.
}\label{tab:1}
\centering
{\footnotesize%
\begin{tabular}{l|cccccccc}\hline
Input & $x_1$ & $x_2$ &$x_3$ &$x_4$ &$x_5$ &$x_6$ &$x_7$ &$x_8$ \\ \hline
First order effect & 0.0033 &  -0.0042 &   0.0010 &   0.0033 &  -0.0020 &   0.0238 &   0.0287  &  0.0284 \\
Shapley (subset) &0.0209  &  0.0327  &  0.0163  &  0.0307  &  0.0105 &  0.0359 &   0.0418  &  0.0560 \\
Shapley (superset) &0.0214 &   0.0313 &   0.0202 &   0.0318 &   0.0146  &  0.0347  &  0.0395 &   0.0596 \\
Total effect & 0.0570  &  0.0626 &   0.0360  &  0.0603  &  0.0222  &  0.0394  &  0.0571  &  0.0866 \\ \hline
Input & $x_9$ &$x_{10}$ &$x_{11}$ &$x_{12}$ &$x_{13}$ &$x_{14}$ &$x_{15}$ & Sum\\ \hline
First order effect & 0.0593 &   0.0090 &   0.1078 &   0.1178 &   0.1149  &  0.1030 &   0.1322 & 0.7262\\
Shapley (subset) & 0.0837 &   0.0219  &  0.1239   & 0.1232  &  0.1353  &  0.1208  &  0.1464& 1.0000\\
Shapley (superset) &0.0762 &   0.0239 &   0.1186  &  0.1420  &  0.1238 &   0.1210  &  0.1412& 1.0000\\
Total effect & 0.1031  &  0.0364  &  0.1517  &  0.1526  &  0.1444 &   0.1429  &  0.1583& 1.3105 \\ \hline
\end{tabular}}
\end{table}
%Reducing foods environmental impacts through
%producers and consumers
%J. Poore* and T. Nemecek
%*Corresponding author. Email: joseph.poore@queens.ox.ac.uk
%Published 1 June 2018, Science 360, 987 (2018)
%

\subsection{Fire-Spread Model} The fire-spread simulator used in \cite{SongNels16} is one of the first realistic 
applications on which algorithms for the estimation of Shapley effects were tested. 
 Other sensitivity analyses of the Rothermel fire-spread model are performed in \cite{LiuJime15}, 
 using a slightly different set of equations than what is discussed here, and in \cite{BallPare18}.
 A state-of-the-art report for this fire-spread simulation model is available as \cite{Andr18}.  
 Starting point of our analysis has been
 the implementation of the firespread model 
 which can be found at \url{https://EunhyeSong.info/}, accessed by the authors on 2019/05/27. 
% \todo[inline]{We need to place an acknowledment to Song}
The simulator output is the rate of fire-spread and is calculated from the series of equations detailed in Appendix \ref{app}. As in \cite{SongNels16}, we consider three distributional scenarios. In the first scenario, model inputs are considered independent (no dependence case). In a second scenario an intermediate level of correlation is introduced between $ m_d $ and $ U $. The rationale is that the windier it is, the dryer the fuel gets. Following the terminology in Song et al. (2016), one calls this the ``weak dependence'' scenario. In a third scenario, stronger correlations are introduced among the inputs (as per Song et al. 2016, ``strong dependence'' scenario). Numerically, the ``weak dependence'' scenario assumes a rank correlation of $-0.3$, the ``strong dependence'' scenario a rank correlation of $-0.8$.

The inputs and their marginal distributions are listed in Table \ref{tab:fire}.
 \begin{table}
 	\caption{Fire-spread example: Description of inputs and their marginal distributions. 
 		Note that SI units from column 3 are used for describing the probability distributions, while the formulas use imperial units from column 4.
 		The strong wind speed scenario input is already included in the distribution of $U$.}\label{tab:fire}
 	\centering
 	{\footnotesize
 		\begin{tabular}{llllll}
 			\hline
 			Variable&Description&Units (D)& Units (E)&Distribution&Trunc.\\
 			\hline
 			$\delta$ &Fuel depth &cm &  ft &  logn(2.19,.517) & $ $\\
 			$\sigma$ &Fuel particle area to volume ratio &cm\textsuperscript{-1}&ft\textsuperscript{-1}&logn(3.31,.294) & $>\frac{3}{0.6}$ \\
 			$h$ &Fuel particle low heat content&kcal/kg&btu/lb&     logn(8.48,.063) & \\
 			$\rho_P$ & Oven-dry particle density&g/cm\textsuperscript{3} & lb/ft\textsuperscript{3} & logn(-.592,.219) & \\
 			$m_l$ & Moisture content of live fuel&g/g& &      norm(1.18,.377) & $>0$ \\
 			$m_d$ & Moisture content of dead fuel&g/g& &    norm(.19,.047) & \\
 			$S_T$ & Fuel particle total mineral content&g/g&  & norm(.049,.011) & $>0$\\
 			$U$ & Wind speed at midflame height &km/h& ft/min&     logn(2.9534,.5569) & \\
 			$\tan\Phi$ & Slope& & &      norm(.38,.186) & $>0$ \\
 			$P$ & Dead fuel loading to total fuel loading& & &      logn(-2.19,.66) & $<1$\\ \hline
 \end{tabular}}\end{table}

\begin{figure}
\centering
\includegraphics[width=0.99\textwidth]{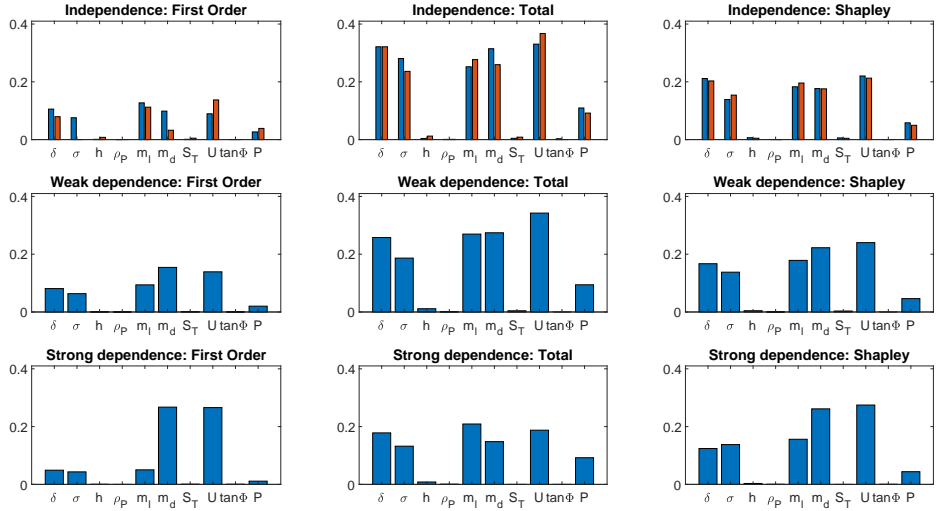}
\caption{Shapley values for the rank-correlated fire-spread model, basic sample size $2^{14}$.}\label{fig:fire}
\end{figure}

We propagate uncertainty in the simulator using a quasi Monte-Carlo generator and a sample size of $2^{14}$. Simulation time is $65.5s $
on the aforementioned machine. 
%\todo[inline]{Elmar, is correct, did we QMC?}
We note that the fire-spread simulator features a mainly multiplicative input-output mapping. 
This, together with the choice of logarithmic distributions make the simulator output span several orders of magnitude. 
In these situations, it has been underlined in previous literature \cite{ImanHora90,BorgTara14} that a slow convergence 
in variance-based estimators can be expected.
Indeed, Figure \ref{fig:fire} shows that the first order effects and the total effects differ
depending upon using the Jansen or the Sobol'-Saltelli estimators, even for the relative large basic sample size of $2^{14}$. 
However, despite the fluctuations in variance-based sensitivity measures, the Shapley effects estimates are relatively robust.
For the dependent input sample case which postulates two physical plausible scenarios, linking the wind speed and the moisture content together, 
the duality of the estimators 
breaks down and we are left with essentially one estimator (the one combining \verb|yi| with \verb|yb|).

Under strong dependence the total effect of $m_d$
becomes lower than the corresponding first order effect.

\begin{figure}
\centering
\includegraphics[width=0.99\textwidth]{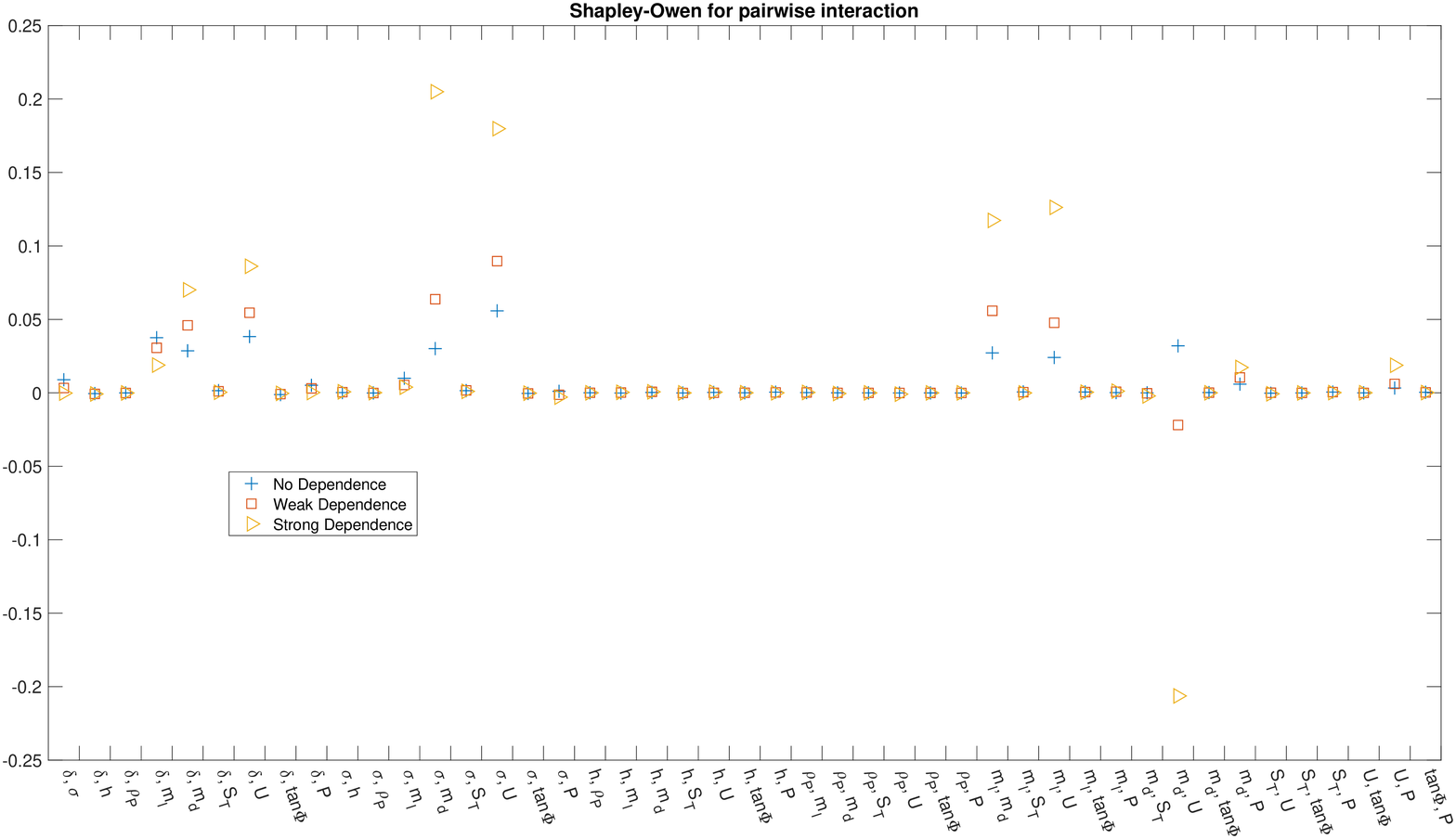}
\caption{Pairwise Shapley-Owen effects for the rank-correlated fire-spread model.}\label{fig:shapO}
\end{figure}

As mentioned, the algorithm enables the computation of Shapley-Owen interaction effects. 
Figure  \ref{fig:shapO} reports the pairwise Shapley-Owen effects for the three above mentioned samples, 
with no dependence ($+$), weak dependence ($\square$) and strong dependence ($ \triangleright $).
Figure  \ref{fig:shapO} shows that Shapley-Owen effects have absolute values varying from 0 to 0.2, with 30 out of 45 effects 
null in all three cases of statistical dependence. In the reminder, we shall focus on interactions for which the magnitude of Shapley-Owen effects is higher than 0.01, calling them significant.
% \todo[inline]{Giov, Elmar: please check that the numbers I inputed are not too far from correct, maybe also using the original numbers from the code?}
%\todo[inline]{Please check if the red part is clear and makes sense!}
Input $ \delta $ is involved in significant interactions with $ m_l $, $ m_d  $ and $ U $. 
Input $ \sigma $ in turn, is involved in significant interactions with $ m_d $ and $ U $. 
We then register significant interactions between $m_l$ and $m_d$, $m_l$ and $U$, and between $ m_d$ and $U$.
Note that $h$, $\rho_P$ and $\tan \Phi$ show neither Shapley nor Shapley-Owen effects. 

As the input correlation increases, a first, overall impression would suggest that the presence of dependence 
tends to increase the explanatory power of interaction effects. For instance, note the increases in the magnitude of $ \phi_{\sigma,m_d} $ (from 0.03 to 0.21), of $ \phi_{\sigma,U} $ (from $0.055$ to $0.18$). This means that the joint explanatory power of $\sigma$ with $m_d$ and with $U $ increases as the correlation among the inputs increases. However, this effect is not systematic. The absolute value of $ \phi_{\delta,m_l} $ decreases from about $ 0.04 $ in the \textquotedblleft no dependence\textquotedblright scenario, to $ 0.02  $ in the \textquotedblleft strong dependence\textquotedblright scenario. The Shapley-Owen effect $ \phi_{m_d,U} $ deserves some attention. First, $ \phi_{m_d,U} $ grows in magnitude from 0.03 to 0.21. However, $ \phi_{m_d,U} $ changes sign as we move from the  no dependence to the weak dependence and to the strong dependence cases. Thus, these two inputs interact negatively (and strongly negatively) as their negative correlation increases, meaning that they lose 
explanatory power when considered together. 
Let us consider the variance-based indices of these two variables. In the no dependence case, their total variance-based indices are greater than they first order indices. In the weak dependence case, the first-order indices are still smaller than the corresponding total ones and the Shapley-Owen effect $ \phi_{m_d,U} $ is now negative but with a small absolute value. In the third case, strong dependence becomes strong, 
this Shapley-Owen interaction is highly negative and we can see that the total effects for $m_d$ and $U$ are 
lower than their corresponding first-order indices. We also observe that in the strong dependence case, for these two inputs, the bracketing property of \cite{Owen14} 
doesn't hold anymore. For instance, we observe the inversion $T_{m_d}\leq \Phi_{m_d}\leq S_{m_d}$. 
The works \cite{SongNels16, IoosPrie19, RabiBorg19x} discuss other examples of inversion of the bracketing 
property when inputs become dependent.
In this respect, Shapley-Owen effects can offer new insights useful to understand the origin of this inversion.

\section{Conclusions}
%\todo[inline]{I have changed the conclusions, please have a look}
Algorithms for computing Shapley values are attracting increasing interest.
In this work, we have proposed an approach based on the M\"obius inverse that reduces computational burden notably. 
The approach enables the computation not only of Shapley effects, but also of Shapley-Owen effects for groups.
The algorithm is exact in the sense
that all possible input combinations are considered and is valid for all dependent input cases that can to be addressed by Rosenblatt transformations. 
%\todo[inline]{Is this dependence the same of Song and Nelson? Is this restrictive?}
In terms of future research, we believe that algorithms based on the M\"obius-inverse representation of Shapley effects 
might be beneficial also in the given-data context, for which currently only the permutation-based 
algorithm of \cite{BrotBach18x} is available. 
We are currently investigating this possibility.
%The algorithms allow for an exchange of the value functions.
\section*{Acknowledgments}
Parts of this work were conducted while the first author was visiting Bocconi University.
The authors thank Eunhye Song for providing the code of the fire-spread simulator. 
\appendix
\section{Equations of the Fire-Spread Simulator Used in this Work}\label{app}
The simulator output is given by
\begin{alignat}{2}
 R&=I_R\cdot\xi\cdot\frac{1+\Psi_W+\Psi_S}{\rho_b\cdot\epsilon\cdot Q_\text{ig}}&& \qquad\text{rate of fire-spread [ft/min]}\\
 \intertext{which is obtained from the following subequations}
\omega_0 &=\frac{0.2048}{1+\exp\left(\frac{15-30.48\delta}{2}\right)}&& \qquad\text{fuel loading [lb/ft\textsuperscript{2}]}\\
\Gamma_{\max}&=\sigma^{1.5}/(495+.0594\sigma^{1.5})&& \qquad\text{maximum reaction velocity [1/min]} \\
\beta_\textrm{op}&=3.348\sigma^{-0.8189}&&                    \qquad\text{optimum packing ratio}\\
%\end{alignat}
%\begin{alignat}{2}
A &=133.0\sigma^{-0.7913}&&             \qquad\text{Albini 1976} \\
\theta_{\ast}&=\frac{301.4-305.87(m_l -m_d) + 2260m_d}{2260m_l}&& \\
\theta&=\min(1,\max(\theta_{\ast},0))&&  \\
%\end{alignat}
%\begin{alignat}{2}
\mu_M&=\!\exp(-7.3P m_d\!-\!(7.3 \theta\!+\!2.13)(1\!-\!P) m_l)&& \qquad\text{moisture damping coefficient}\\
\mu_S&=0.174S_T^{-0.19}&&                       \qquad\text{mineral damping coefficient}\\
C &=7.47\exp(-0.133\sigma^{0.55})&&\\
B &=0.02526\sigma^{0.54}&&\\
E &=0.715\exp(-3.59\cdot 10^{-4}\sigma)&& \\
%\end{alignat}
%\begin{alignat}{2}
\omega_n&= \omega_0(1-S_T)&&                          \qquad\text{net fuel loading [lb/ft\textsuperscript{2}]}\\
\rho_b&= \frac{\omega_0}{\delta}&&                            \qquad\text{ovendry bulk density [lb/ft\textsuperscript{3}]}\\
\epsilon&=\exp(-138/\sigma)&&                        \qquad\text{effective heating number}\\
Q_\text{ig}&=130.87+1054.43m_d&&                     \qquad\text{heat of preignition [Btu/lb]}\\
%Qig &= (401.41 + m_d * 2565.87) * 0.4536/1.060&&\\
\beta&=\frac{\rho_b}{\rho_P}&&                          \qquad\text{packing ratio} 
% Gamma&=\Gamma_\max.*beta(x)./ \beta_\textrm{op}.*A(x).*exp(A(x).*...
%     (1-beta(x)./ \beta_\textrm{op}))&&                       \qquad\text{ optimum reaction velocity
\end{alignat}
\begin{alignat}{2}
\Gamma&=\Gamma_{\max}\left(\frac{\beta}{\beta_\textrm{op}}\right)^{A} \exp\left(A
\left(1-\frac{\beta}{\beta_\textrm{op}}\right)\right)&&                       \qquad\text{optimum reaction velocity [1/min]} \\
\xi &=\frac{\exp\left((0.792 + 0.681\sqrt{\sigma})(\beta+ 0.1)\right)}{
	192 + 0.2595\sigma}&&                         \qquad\text{propagating flux ratio} \\
\Psi_W&=C U^B
\left(\frac{\beta}{\beta_\textrm{op}}\right)^{-E}&&                   \qquad\text{wind coefficient}
\end{alignat}
\begin{alignat}{2}
\Psi_S&=5.275\beta^{-0.3}(\tan\Phi)^2&&         \qquad\text{slope factor}\\
I_R&=\Gamma\cdot\omega_n\cdot h \cdot \mu_M \cdot \mu_S&&    \qquad\text{reaction intensity [Btu/ft\textsuperscript{2}min]}
\end{alignat}
The corresponding Matlab implementation is available opon request. %as online supplement.
%\todo[inline]{Elmar, please insert the Matlab code}
%
%\bibliographystyle{siamplain}
%\bibliography{../Elmar/TeXStuff/sensitivity}

\end{document}